\documentstyle[twoside,fleqn,espcrc2,epsf]{article}

\title{Quantum chaos in QCD at finite temperature}

\author{H.~Markum$^{\rm a}$, 
  R.~Pullirsch$^{\rm a}$\thanks{Poster presented by R.~Pullirsch}, 
  K.~Rabitsch\address{Institut f\"ur Kernphysik, Technische
  Universit\"at Wien, A-1040 Wien, Austria}, and 
  T.~Wettig\address{Institut f\"ur Theoretische Physik, 
  Technische Universit\"at M\"unchen, D-85747 Garching, Germany}}
       
\begin{document}

\begin{abstract}
  We study complete eigenvalue spectra of the staggered Dirac matrix
  in quenched QCD on a $6^3\times 4$ lattice.  In particular, we
  investigate the nearest-neighbor spacing distribution $P(s)$ for
  various values of $\beta$ both in the confinement and deconfinement
  phase.  In both phases except far into the deconfinement region, the
  data agree with the Wigner surmise of random matrix theory which is
  indicative of quantum chaos.  No signs of a transition to Poisson
  regularity are found, and the reasons for this result are discussed.
\end{abstract}

\maketitle

\section{Introduction}

The fluctuation properties of the eigenvalues of the lattice Dirac
operator have attracted much attention recently.  In
Ref.~\cite{Hala95}, it was first demonstrated that they are described
by random matrix theory (RMT).  In particular, the so-called
nearest-neighbor spacing distribution $P(s)$, i.e., the distribution
of spacings $s$ between adjacent eigenvalues, agrees with the Wigner
surmise of RMT.  According to a conjecture by Bohigas \cite{Bohi84},
quantum systems whose classical counterparts are chaotic have a $P(s)$
given by RMT whereas systems whose classical counterparts are
integrable obey a Poisson distribution, $P(s)=e^{-s}$.  In this sense,
the form of $P(s)$ characterizes quantum chaos.

At the localization transition in disordered mesoscopic systems, one
observes a transition in $P(s)$ from Wigner to Poisson behavior.  The
question of whether there is such a transition in the case of the
lattice Dirac operator was first raised in Ref.~\cite{Hala95}.  The
present study serves to investigate this question.  Recently, a Wigner
to Poisson transition was also studied in the context of a spatially
homogeneous Yang-Mills-Higgs system \cite{Sala97}.

The lattice Dirac operator falls into one of three symmetry classes
corresponding to the three chiral Gaussian ensembles of RMT
\cite{Verb94}.  In Ref.~\cite{Hala95}, the Dirac matrix was studied
for SU(2) using both staggered and Wilson fermions which correspond to
the chiral Gaussian symplectic (chGSE) and orthogonal (chGOE)
ensemble, respectively.  Here, we study SU(3) which for both staggered
and Wilson fermions corresponds to the chiral Gaussian unitary
ensemble (chGUE).  We thus cover the last remaining symmetry class.
The RMT result for $P(s)$ is quite complicated; it can be expressed in
terms of so-called prolate spheroidal functions, see
Ref.~\cite{Meht91} where $P(s)$ has also been tabulated.  A very good
approximation to $P(s)$ is provided by
\begin{equation}
  \label{eq1}
  P(s)=\frac{32}{\pi^2}s^2e^{-\frac{4}{\pi}s^2}
\end{equation}
which is the Wigner surmise for the chGUE.

\section{Eigenvalue analysis}

We generated gauge field configurations using the standard Wilson
plaquette action for SU(3) and constructed the matrix of the Dirac
operator using the Kogut-Susskind prescription.  The Dirac matrix is
anti-hermitian so that all eigenvalues are imaginary and occur in
pairs with opposite sign.  We have worked on a $6^3\times 4$ lattice
with various values of $\beta$ and typically produced 10 independent
configurations for each $\beta$.  All spectra were checked against the
analytical sum rules
\begin{equation}
\sum_{\lambda_n} \lambda_n = 0 \qquad {\rm and} \qquad
\sum_{\lambda_n>0} \lambda_n^2 = 3V \:,
\end{equation}
where V is the lattice volume. 

To construct $P(s)$, one first has to ``unfold'' the spectra
\cite{Bohi84}.  This procedure is a $local$ rescaling of the energy 
scale so that the mean level spacing is equal to unity on the unfolded
scale.  Ensemble and spectral averages (the latter is possible because
of the spectral ergodicity property of RMT) are only meaningful after
unfolding.

\section{Results}

\begin{figure*}[t]
\begin{center}
\begin{tabular}{cccc}
 \phantom{ww} &Confinement: $\beta=2.8$ (solid line) & \phantom{ww}&
 Deconfinement:  $\beta=6.0$  (solid line)\\[0mm]
 & and $\beta=5.0$ (dashed line) & & and $\beta=10.0$ (dashed line)\\[3mm]
 &\epsfxsize=4.9cm\epsffile{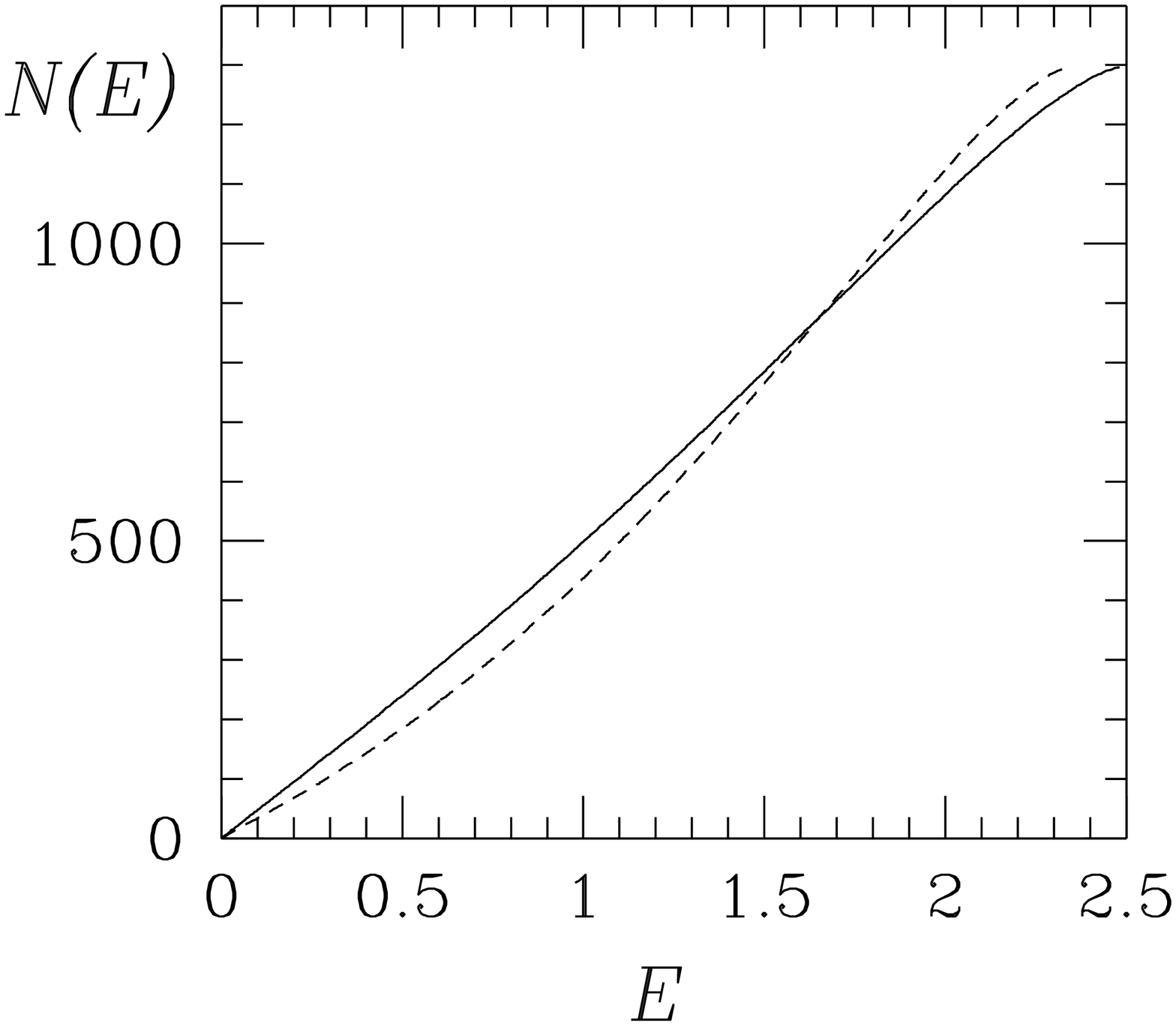} \phantom{W} & &
\epsfxsize=4.9cm\epsffile{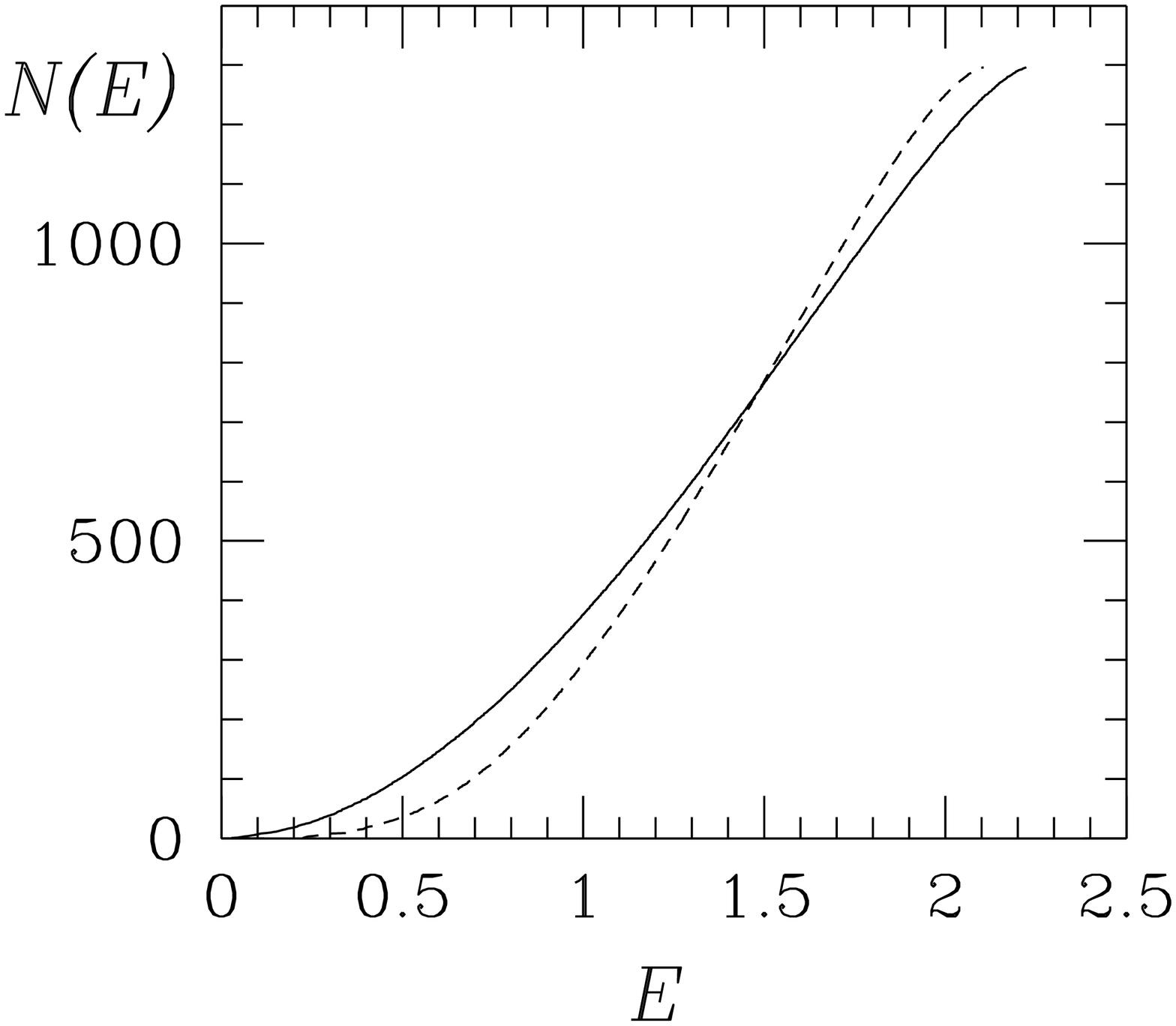} \phantom{W}
\end{tabular}
\end{center}
\vspace*{-12mm}
\caption{The staircase function $N(E)$ representing the number of
  positive eigenvalues $\le E$ for typical configurations on a $6^3
  \times 4$ lattice.}
\label{fig1}
\end{figure*}

Our results are displayed in the plots.  Figure~\ref{fig1} shows the
staircase function $N(E)$ which is defined as the number of
eigenvalues with $\lambda \le E$.  In the continuum,
$N(E)=\int_0^E\rho(\lambda)d\lambda$, where $\rho(\lambda)$ is the
spectral density.  Note that $\rho(\lambda)$ cannot be obtained from
a random-matrix model.

\begin{figure*}[b]
\begin{center}
\begin{tabular}{ccc}
 \phantom{WWW} Confinement: $\beta=2.8$ & \phantom{WWW} &
 \phantom{WWW} Confinement:  $\beta=5.0$\\[3mm] 
\epsfxsize=4.9cm\epsffile{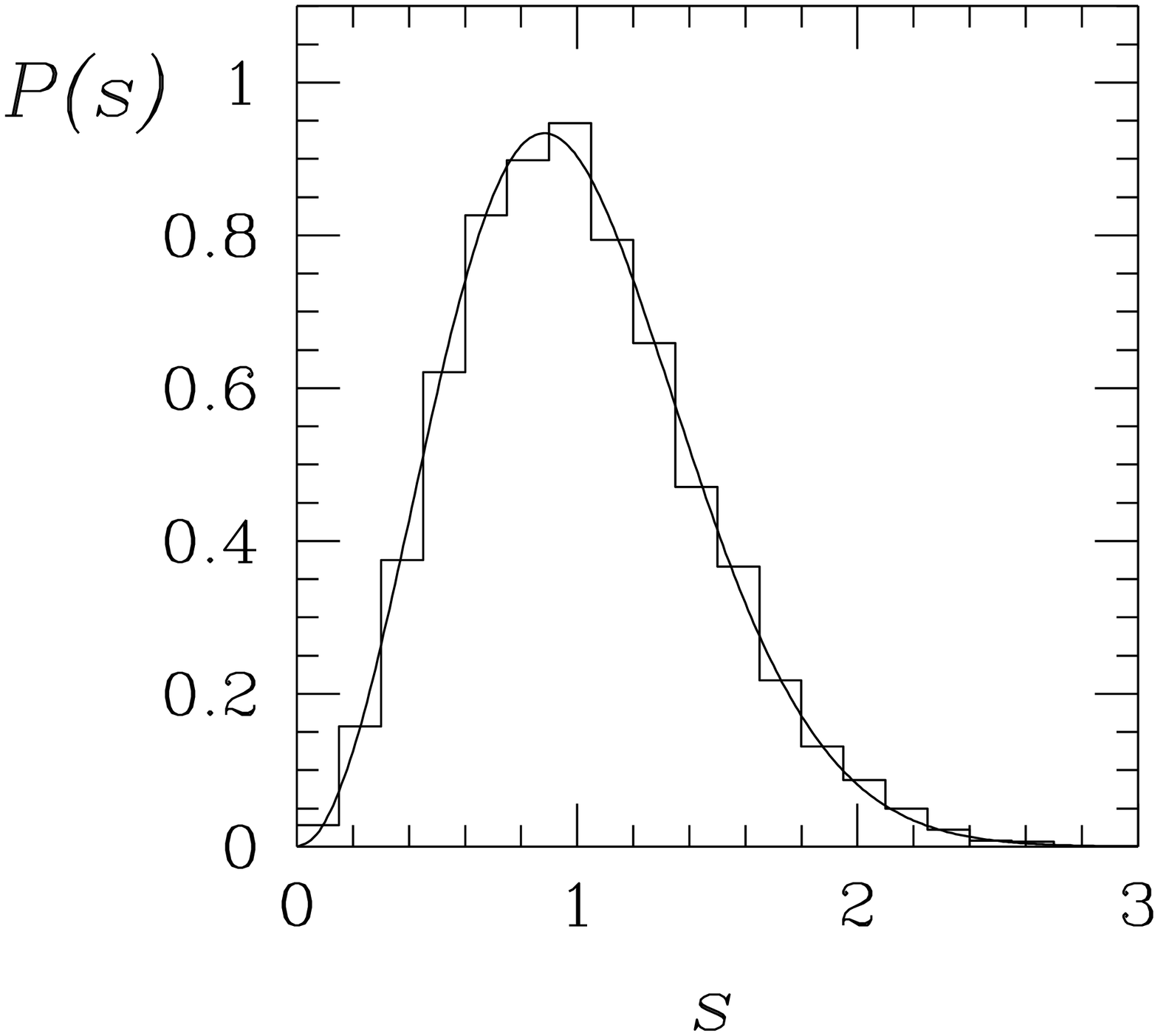} & &
\epsfxsize=4.9cm\epsffile{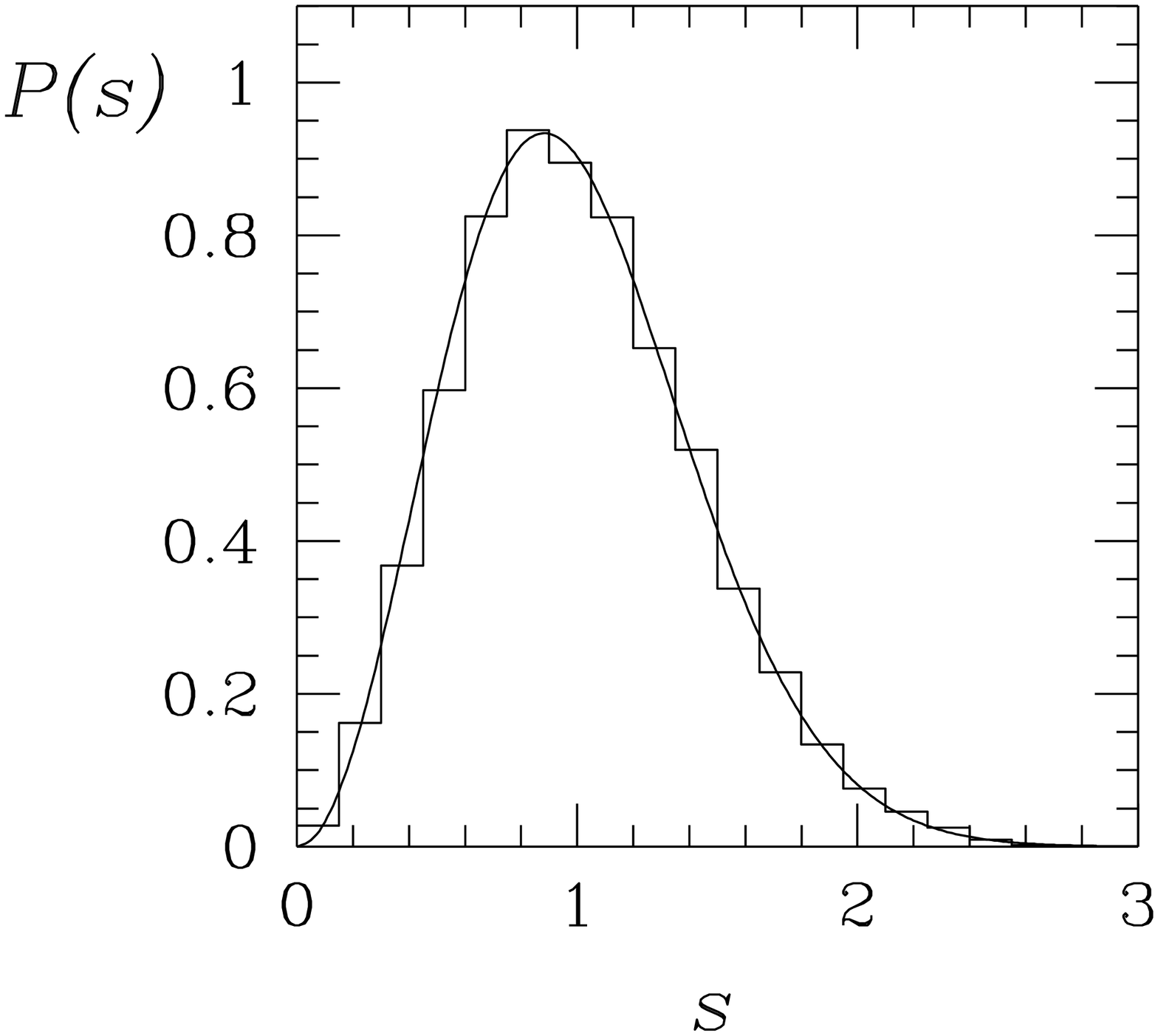}\\[2mm]
 \phantom{WWw} Deconfinement: $\beta=6.0$ & & \phantom{WWw} Deconfinement:
 $\beta=10.0$ \\[3mm]
\epsfxsize=4.9cm\epsffile{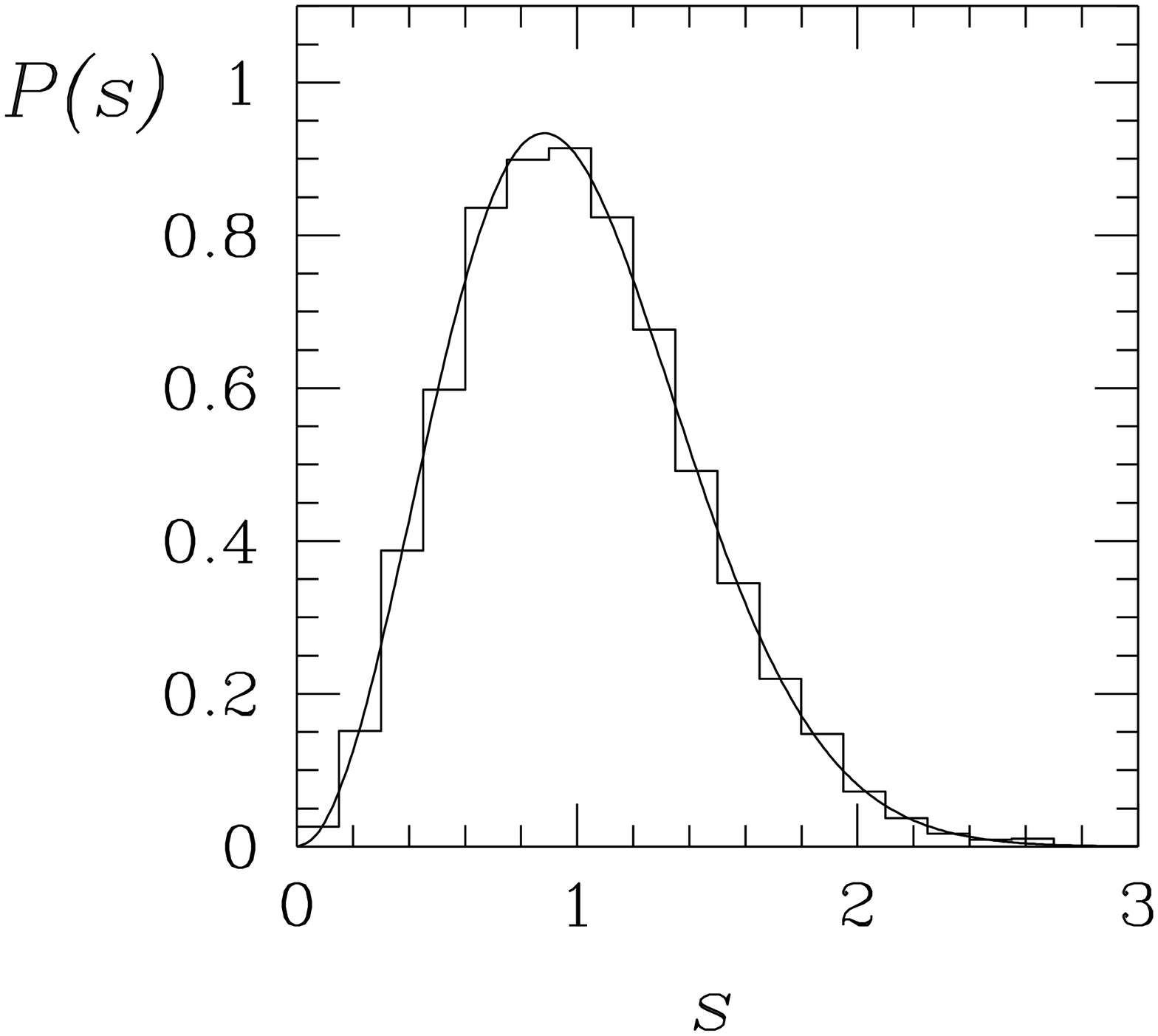} & &
\epsfxsize=4.9cm\epsffile{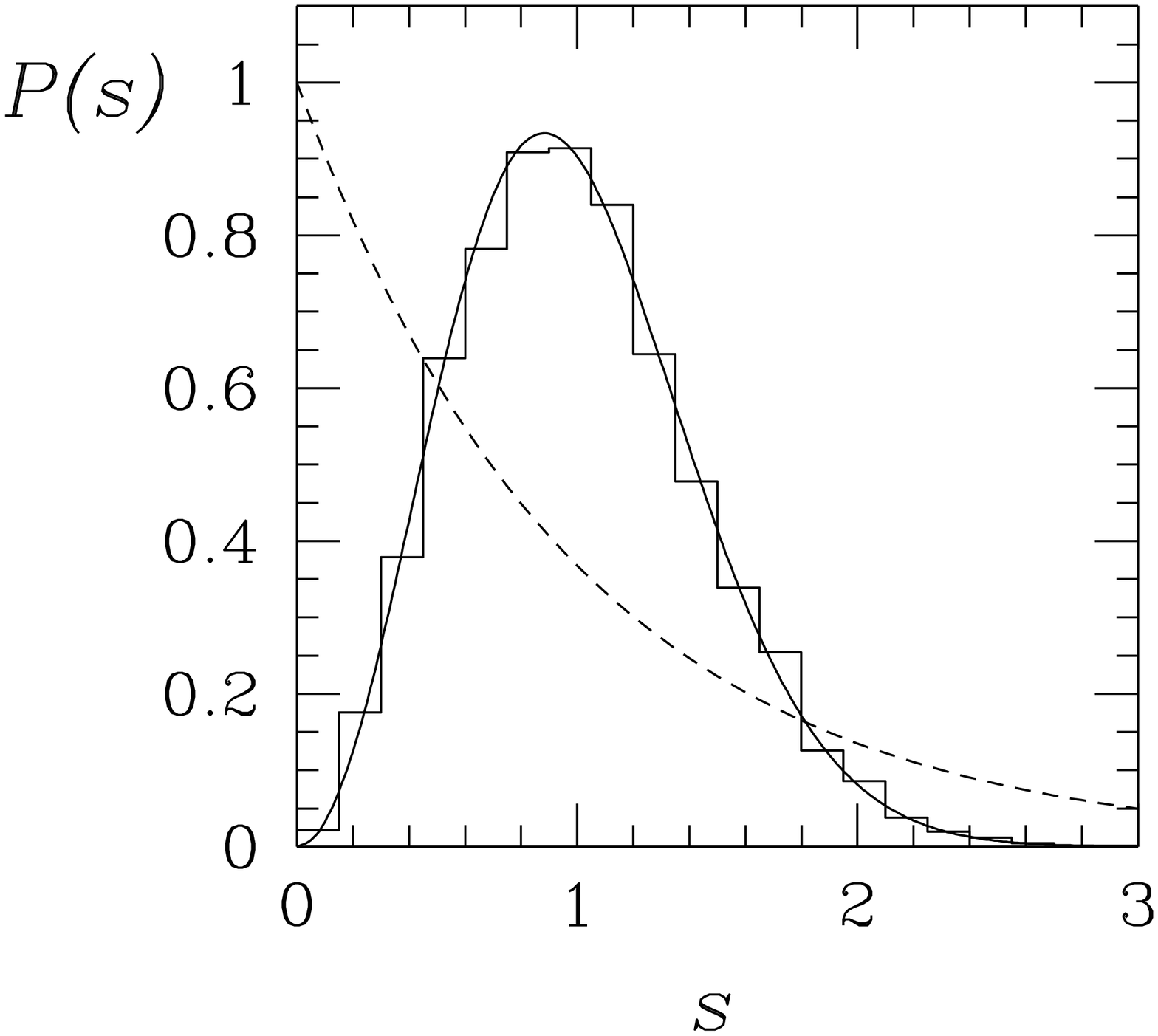}
\end{tabular}
\end{center}
\vspace*{-13mm}
\caption{The nearest-neighbor spacing distribution $P(s)$ averaged
  over 10 independent configurations on a $6^3 \times 4$ lattice
  (histograms) compared with the random-matrix result (solid lines).
  For comparison, the Poisson distribution $P(s)=e^{-s}$ is inserted
  for $\beta=10.0$ (dashed line).  There are no changes in $P(s)$
  across the deconfinement phase transition.}
\label{fig2}
\end{figure*}

Figure~\ref{fig2} shows the nearest-neighbor spacing distribution
$P(s)$ compared with the RMT result.  In the confinement phase, we
find the expected agreement of $P(s)$ with the Wigner surmise of
Eq.~(\ref{eq1}).  In the deconfinement phase, we still observe
agreement with the RMT result up to $\beta=10.0$.  No signs for a
transition to Poisson behavior are found.  Thus, the deconfinement
phase transition does not seem to coincide with a transition in the
spacing distribution.  For larger values of $\beta$ the eigenvalues
start to approach the degenerate eigenvalues of the free theory, given
by $\lambda^2=\sum_{\mu=1}^4 \sin^2(2\pi n_\mu/L_\mu)/a^2$, where $a$
is the lattice constant and $n_\mu=0,\ldots,L_\mu-1$.  In this case,
the nearest-neighbor spacing distribution is trivial.  However, it is
possible to lift the degeneracies of the free eigenvalues using an
asymmetric lattice where $L_x$, $L_y$, etc. are relative primes.  For
large lattices, the nearest-neighbor spacing distribution of the
non-degenerate free eigenvalues is then given by the Poisson
distribution.  While it may be interesting to search for a Wigner to
Poisson transition on such asymmetric lattices, it seems clear that
this transition will not coincide with the deconfinement phase
transition.

Furthermore, we do not think that the absence of a signature for a
transition from Wigner to Poisson behavior at the deconfinement phase
transition is due to the finite lattice size.  Even for the small
lattice size we used, the agreement of $P(s)$ with the RMT curve is
nearly perfect.  This leads us to believe that we should have seen
some sign of a transition if it existed in the thermodynamic limit.


\section{Conclusions}
 
We have searched for a transition in the near\-est-neighbor spacing
distribution $P(s)$ from Wig\-ner to Poisson behavior across the
deconfinement phase transition.  Such a transition exists, e.g., at
the localization transition in disordered mesoscopic systems. In a
Yang-Mills-Higgs system a smooth transition along a Brody distribution
was seen \cite{Sala97}.  We found no signature of a transition in our
lattice data.  The data agree with the RMT result in both the
confinement and the deconfinement phase except for extremely large
values of $\beta$ where the eigenvalues are trivial.

In hindsight, these results are not totally unexpected.  Temporal
monopole currents survive the deconfinement phase transition leading
to confinement of spatial Wilson loops.  Thus, even in the
deconfinement phase, the gauge fields retain a certain degree of
randomness.  In future investigations one might try to disentangle
those spatial contributions to the Dirac matrix.

\section{Acknowledgments}

This work was supported in part by FWF project P10468-PHY.  We thank
E.-M.\ Ilgenfritz, M.I.\ Polikarpov, and J.J.M.\ Ver\-baar\-schot for
helpful discussions.


\end{document}